\documentclass[a4paper]{article}

\usepackage{RR}
\RRNo{6180}
\usepackage{hyperref}

\usepackage{graphicx}
\usepackage{latexsym}
\usepackage{verbatim}

\usepackage{amsmath}
\usepackage{amsfonts}
\usepackage{amssymb}

\usepackage{amsmath, amsthm, amssymb}

\usepackage{epsfig}
\usepackage[dutch,english]{babel}
\usepackage{mathrsfs}

\newtheorem{Theorem}{Theorem}

\newtheorem{Lemma}[Theorem]{Lemma}

\newtheorem{Corollary}[Theorem]{Corollary}

\RRdate{April 2007}
\RRauthor{
Natalia Osipova
\thanks{PhD student at INRIA Sophia Antipolis, France, e-mail: Natalia.Osipova@sophia.inria.fr
} \thanks{{The work was supported by France Telecom R\&D Grant
``Mod\'elisation et Gestion du Trafic R\'eseaux Internet'' no.
42937433.}}
}

\authorhead{N. Osipova}
\RRtitle{La File d'Attente Avec Service à Temps Partagé Avec
Arrivées en Rafale }

\RRetitle{Batch Processor Sharing with\\ Hyper-Exponential Service
Time }
\titlehead{Batch Processor Sharing with Hyper-Exponential Service
Time }

\RRresume{ Nous étudions la file d'attente avec service à temps
partagé avec arrivées en rafale, BPS (``Batch Processor
Sharing''), lorsque les temps de service ont une distribution
hyper-exponentielle et que le processus d'arrivée des rafales est
Poisson. Une des motivations de cette étude est que la file BPS
apparaît naturellement lorsque l'on étudie les procédures
d'ordonnancement favorisant les connexions courtes sur un réseau
internet. Dans le cas d'une distribution hyper-exponentielle des
temps de service nous trouvons une expression analytique pour le
temps de réponse moyen conditionné sur le temps de service pour
une file d'attente BPS. Nous montrons que ce temps de réponse
moyen conditionné est une fonction concave du temps de service.
Nous appliquons les résultats obtenus à la file d'attente munie
d'une politique à Temps Partagées avec Deux-Niveaux ``Two Level
Processor Sharing'' (TLPS) avec distribution hyper-exponentielle
des temps de service. Nous trouvons l'expression du temps de
réponse moyen pour la file TLPS. La discipline de service TLPS
peut être utilisée pour ordonner l'accès aux ressources en
fonction de la taille dans un réseau TCP/IP ou sur un serveur
Web.}

\RRabstract{ We study Processor-Sharing queueing model with the
hyper-exponential service time distribution and Poisson batch
arrival process. One of the main goals to study Batch
Processor-Sharing (BPS) is the possibility of its application to
size-based scheduling, which is used in differentiation between
short and long flows in the Internet. In the case of
hyper-exponential service time distribution we find an analytical
expression for the expected conditional response time for the BPS
queue. We show that the expected conditional response time is a
concave function of the service time. We apply the obtained
results to the Two Level Processor-Sharing (TLPS) model with the
hyper-exponential service time distribution and find an expression
of the expected response time for the TLPS model. The TLPS
scheduling discipline can be applied to size-based differentiation
in TCP/IP networks and Web server request handling. }

\RRmotcle{La file d'attente avec service à temps partagé avec
arrivées en rafale, la file d'attente munie d'une politique à
Temps Partagées avec Deux-Niveaux, réseau TCP/IP, distribution
hyper-exponentielle, Laplace transforment, Matrices de Cauchy. }

\RRkeyword{ Batch Processor Sharing, Two Level Processor sharing,
Hyper-Exponential distribution, Laplace transform, Cauchy
matrices.}

\RRprojet{MAESTRO}
\RRtheme{\THCom}
\URSophia 

\begin{document}

\makeRR

\newpage
\section{Introduction}
\mbox{}\\
The Processor-Sharing (PS) queueing systems are now often used to
model communication and computer systems. The PS systems were
first introduced by Kleinrock (see {\cite{kleinrock} and
references therein}). Under the PS policy each job receives an
equal share of the processor.
\\\\
PS with batch arrivals (BPS) is not yet characterized fully.
Kleinrock \emph{et al.} {\cite{Kle_1}} first studied BPS. They
found that the derivative of the expected response time satisfies
the integral equation and found the analytical solution in the
case when the job size (service time) distribution function has
the form \mbox{$F(x)=1-p(x) e^{-\mu x}$} where $p(x)$ is a
polynomial.
\\\\
Bansal {\cite{Bansal}}, using Kleinrock's integral equation,
obtained the solution for the Laplace transform of the expected
conditional service time as a solution of the system of linear
equations, when the job size distribution is a hyper-exponential
distribution. Also he considers distributions with a rational
Laplace transform. Rege and Sengupta {\cite{Rege_Singupta}}
obtained the expression for the response time in condition upon
the number of customers in the system. Feng and Mishra
{\cite{Feng_Misra}} provided bounds for the expected conditional
response time, the bounds depend on the second moment of the
service time distribution. Avrachenkov \emph{et al.}
{\cite{AAB_BPS_MLPS_QS}} proved existence and uniqueness of the
solution of the Kleinrock's integral equation and provided
asymptotic analysis and bounds on the expected conditional
response time.
\\\\
We study the BPS model with the hyper-exponential service time
distribution. For this distribution we provide the solution of the
Kleinrock's integral equation according to the derivative of the
expected conditional response time. We prove the concavity of the
expected conditional sojourn time function for the BPS model with
the hyper-exponential job size distribution function. We note that
the concavity of the expected conditional sojourn time for the BPS
with the hyper-exponential job size distribution was proven by
another method in {\cite{Kim_and_Kim}}.
\\\\
One of the main goals to study BPS is the possibility of its
application to age-based scheduling and the possibility to take
into account the burstiness of the arrival process. Bursty
arrivals often occur in such modern systems as web server.
Age-based scheduling is used in differentiation of short and long
flows in the Internet. A quite general set of age-based scheduling
mechanisms was introduced by Kleinrock and termed as Multy-Level
PS (MLPS).
\\\\
In MLPS jobs are classified into different classes depending on
their attained amount of service. Jobs within the same class are
served according to FCFS (First Come First Serve), PS (Processor
Sharing) or FB (Foreground Background) policy. The classes
themselves are served according to the FB policy, so that the
priority is given to the jobs with small sizes.
\\\\
We study the particular case of MLPS, Two-Level PS (TLPS)
scheduling mechanism. It is based on the differentiation of jobs
according to some threshold and gives priority to jobs with small
sizes. The TLPS scheduling mechanism could be used to model such
applications as size based differentiation in TCP/IP networks and
Web server request differentiation.
\\\\
It was shown in {\cite{Aalto_Ayesta}} that when a job size
distribution has a decreasing hazard rate, then with the selection
of the threshold the expected sojourn time of the TLPS system
could be reduced in comparison to standard PS system.
\\\\
The distribution of file sizes in the Internet has a decreasing
hazard rate and often modelled with heavy-tailed distributions. In
{\cite{feldmann_whitt}} it is shown that heavy-tailed
distributions such as Pareto distribution could be approximated
with the hyper-exponential distributions with a significant number
of phases.
\\\\
Therefore, we study the TLPS model with the hyper-exponential
service time distribution. We apply the results of the BPS
queueing model to the TLPS model and find an expression of
expected conditional sojourn time for the TLPS model.
\\\\
The paper is organized as follows. In Section {\ref{sec_BPS}} the
BPS scheduling mechanism is considered in the case when the job
service time distribution is the hyper-exponential distribution.
For this case we provide the solution of the Kleinrock's integral
equation. In Section {\ref{sec_TLPS}} the results obtained for the
BPS model are applied to the TLPS model, where the job size
distribution is also hyper-exponential. An analytical expression
of the expected sojourn time is found. We put some technical
proofs in the Appendix.


\newpage
\section{The Analysis of the Batch Arrival Processor Sharing Queue}
\label{sec_BPS}
\subsection{Main definitions}
\mbox{}\\
Let us consider an M/G/1 system with batch arrivals and
Processor-Sharing (PS) queueing discipline. The batches arrive
according to a Poisson process with arrival rate $\lambda$. Let
$\overline{n}>0$ be the average size of a batch. Let $b>0$ be the
average number of jobs that arrive with (and in addition to) the
tagged job.
\\\\
Let $B(x)$ be the required job size (service time) distribution
and $\overline{B}(x)=1-B(x)$ be its complementary distribution
function.
\\\\
The load is given by $\rho=\lambda \overline{n} m$, with
$m=\int_{0}^{\infty}{x dB(x)}$. We consider that the system is
stable, $\rho<1$.
\\\\
Let $\alpha(x)$ be the expected conditional response time for a
job with service time $x$ and $\alpha'(x)$ be its derivative.
Kleinrock showed in {\cite{kleinrock}} that $\alpha'(x)$ satisfies
the following integro-differential equation
\begin{eqnarray}\label{IntAlpha}
&& \alpha'(x)=\lambda \overline{n} \int_{0}^{\infty} \alpha'(y)
\overline{B}(x+y)dy + \lambda \overline{n} \int_{0}^{x} \alpha'(y)
\overline{B}(x-y)dy +b \overline{B}(x)+1.
\end{eqnarray}
\\
We are interested in the case when the job size distribution is a
hyper-exponential function
\begin{eqnarray}{\label{eq:B(x)}}
&& B(x)=1-\sum_{i=1}^{N}{p_i \,e^{-\mu_i\,x}},\,\,
{\sum_{i}{p_i}=1},\,\, p_i>0,\, \mu_i>0, \quad i=1,...,N, \quad
1<N \leq \infty.
\end{eqnarray}
By $\sum_{i}$ and $\prod_{i}$ we mean $\sum_{i=1}^{N}$ and
$\prod_{i=1}^{N}$. By $\sum_{i\neq j}$ or $\prod_{i\neq j}$ we
mean \mbox{$\sum_{i=1,...,N,\, i\neq j}$} and $\prod_{i=1,...,N,\,
i \neq j}$. By $\forall i$ we mean $i=1,...,N$.
\\\\
Without loss of generality, we can assume that
\begin{eqnarray}{\label{eq:mu_cond}}
&& 0<\mu_N<\mu_{N-1}<...<\mu_2<\mu_1<\infty.
\end{eqnarray}

\subsection{The expected conditional sojourn time for the BPS model}
\mbox{} \\Let us first prove additional Lemma.
\begin{Lemma} {\label{lemma_Psi_solut}} The zeros $b_i$ of the
rational function
\begin{eqnarray}{\label{eq:b_i_find_eq}}
&& \Psi(s)=^{def}1-\lambda \overline{n} \sum_{i}{\frac { p_i
}{s+\mu_i}} =\frac{\prod_{i}(s+b_i)}{\prod_{i}(s+\mu_i)}
\end{eqnarray}
are all real, distinct, positive and satisfy the following
inequalities:
\begin{eqnarray}{\label{eq:b_i_cond}}
0<b_N<\mu_N,\,\,\,\mu_{i+1}<b_i<\mu_i,\quad i=1,...,N-1.
\end{eqnarray}
\end{Lemma}
\begin{proof}[\emph{\textbf{Proof.}}]
Let us study the following equation
$$\Psi(s)=0.$$
Following the approach of \cite{Fayolle_1} we get that it has
$N_1$ roots \mbox{$-b_i,\, i=1,...,N_1$}, where $N_1$ is the
number of distinct elements within $\mu_i$. As \mbox{$\mu_i,\,
i=1,...,N$} are all different, \mbox{$\mu_i \neq \mu_j, \, \,i
\neq j$}, then \mbox{$N_1=N$} and there are $N$ different roots
$-b_i$. All $-b_i$ are real, distinct, negative, satisfy the
following inequalities: \mbox{$0>-b_N>-\mu_N$,
$-\mu_{i+1}>-b_i>-\mu_i,\quad i=1,...,N-1$}.
With this we prove the statement of Lemma~\ref{lemma_Psi_solut}.
So, then it is possible to present $\Psi(s)$ in the following
form:
$$\Psi(s)=\frac{\prod_{i}(s+b_i)}{\prod_{i}(s+\mu_i)}, $$ and from
here we have (\ref{eq:b_i_find_eq}).
\end{proof}
\mbox{}\\
Before presenting our main result let us prove an
auxiliary Lemma.
\begin{Lemma} {\label{lemma_Cauchy_sys}}
The solution of the following system of linear equations:
\begin{equation}{\label{eq:x_j_sys}}
\begin{array}{l}
\displaystyle \sum_{j}\frac{x_j }{\mu_q^2-b_j^2}= 1,\quad
q=1,...,N,
\end{array}
\end{equation} is given by
\begin{eqnarray}{\label{x_k}}
x_k= \frac {\displaystyle\prod_{q=1,...,N} {(\mu_q^2-b_k^2)}} {
\displaystyle\prod_{q\neq k}{(b_q^2-b_k^2)} }, \quad  k=1,...,N.
\end{eqnarray}
\end{Lemma}
\begin{proof}[\emph{\textbf{Proof.}}] The proof is given in the appendix.
\end{proof}
\begin{Corollary} {\label{crlr:x_k_positive}} The solution of equation
({\ref{eq:x_j_sys}}) is positive. Namely, $x_k>0$ for $k=1,...,N$.
\end{Corollary}
\begin{proof}[\emph{\textbf{Proof.}}] As ({\ref{eq:mu_cond}}) and ({\ref{eq:b_i_cond}}), then
the statement of the Corollary holds.
\end{proof}
\mbox{}\\
Now we can prove our main result.
\begin{Theorem}{\label{th_alpha_result}} The
expected conditional response time for BPS queue with
hyper-exponential job size distribution function as in
(\ref{eq:B(x)}) is given by:
\begin{eqnarray}{\label{eq:alpha(x)}}
&&\alpha(x)=c_0 x - \sum_k{\frac{c_k}{b_k}e^{-b_k
x}}+\sum_k\frac{c_k}{b_k}, \quad \alpha(0)=0,
\end{eqnarray}
where
\begin{eqnarray*}
&& c_0=\frac{1}{1-\rho},\\
&& c_k=\frac{b}{2 \lambda \overline{n}}
\left(\frac{\displaystyle\prod_{q} {(\mu_q^2-b_k^2)}} { b_k
\displaystyle\prod_{q\neq k}{(b_q^2-b_k^2)} }\right),
\end{eqnarray*}
and where $b_k$ are the solutions of the following equation:
$$1- \lambda \overline{n}\sum_{i}{\frac { p_i }{s+\mu_i}}=0, $$
and are all positive, distinct, real, satisfy the following
inequalities
$$ 0<b_N<\mu_N,\,\,\,\mu_{i+1}<b_i<\mu_i,\quad i=1,...,N-1. $$
\end{Theorem}
\begin{proof}[\emph{\textbf{Proof.}}] We can rewrite integral equation
(\ref{IntAlpha}) in the following way:
\begin{eqnarray*}
&&{\alpha'(x)=\lambda \overline{n} \int_{0}^{\infty} \alpha'(y)
\sum_{i} p_i  e^{-\mu_i ( x+y)}dy + \lambda \overline{n}
\int_{0}^{x} \alpha'(y) \overline{B}(x-y)dy +b
\overline{B}(x)+1}, \\
&& {\alpha'(x)=\lambda \overline{n} \sum_{i} p_i e^{-\mu_i x}
\int_{0}^{\infty} \alpha'(y) e^{-\mu_i y}dy + \lambda \overline{n}
\int_{0}^{x} \alpha'(y) \overline{B}(x-y)dy +b \overline{B}(x)+1}.
\end{eqnarray*}
We note that in the latter equation \mbox{$\int_{0}^{\infty}
\alpha'(y){e^{-\mu_i y}} dy$}, \,\, \mbox{$i=1,...,N$} are the
Laplace transforms of $\alpha'(y)$ evaluated at \mbox{$\mu_i,\,\,
i=1,...,N$}. Denote
\begin{eqnarray*}
&& L_i=\int_{0}^{\infty} \alpha'(y){e^{-\mu_i y} dy },\quad
i=1,...,N.
\end{eqnarray*}
Then, we have
\begin{eqnarray*}
&& \alpha'(x)=\lambda \overline{n}\sum_{i} p_i L_i e^{-\mu_i x}+{
\lambda \overline{n} \int_{0}^{x} \alpha'(y){\overline{B}( x-y)}dy
+b \overline{B}(x)+1}.
\end{eqnarray*}
Now taking the
Laplace transform of the above equation and using the convolution
property, we get the following equation. Here we denote
${L_{\alpha'}(s)}$ the Laplace transform of $\alpha'(x)$.
\begin{eqnarray*}
&& {L_{\alpha'}(s)}= \lambda \overline{n}\sum_{i}{\frac {{ p_i}
L_i}{s+\mu_i}}+\lambda \overline{n} \sum_{i}{\frac {{ p_i}
L_{\alpha'}(s)}{s+\mu_i}}+{b}\sum_{i}{\frac {p_i} {s+\mu_i}}+
{\frac 1 s},\\
\Longrightarrow && L_{\alpha'}(s) \left (1-\lambda \overline{n}
\sum_{i}{\frac { p_i }{s+\mu_i}}\right)=\lambda \overline{n}
\sum_{i}{\frac {p_i L_i}{s+\mu_i}}+ {b}\sum_{i}{\frac {p_i}
{s+\mu_i}}+ {\frac 1 s}.
\end{eqnarray*}
Let us note that $L_{\alpha'}(\mu_i)=L_i,\quad i=1,...,N$.
According to Lemma~{\ref{lemma_Psi_solut}} and
(\ref{eq:b_i_find_eq}), the following equation holds:
\begin{eqnarray}
&& L_{\alpha'}(s)\frac{\prod_{i}(s+b_i)}{\prod_{i}(s+\mu_i)}
=\lambda \overline{n}\sum_{i}{\frac {p_i L_i}{s+\mu_i}}+
{b}\sum_{i}{\frac {p_i}{s+\mu_i}}+ {\frac 1 s},
\label{eq:L_alpha_diff_mult}\\
\Longrightarrow && L_{\alpha'}(s) =\lambda \overline{n}
\sum_{i}{p_i L_i {\frac{\prod_{k \neq
i}(s+\mu_k)}{\prod_{i}(s+b_i)}}}+ {b} \sum_{i}{p_i}{\frac{\prod_{k
\neq i}(s+\mu_k)}{\prod_{i}(s+b_i)}}+ {\frac 1
s}{\frac{\prod_{i}(s+\mu_i)}{\prod_{i}(s+b_i)}}.{\label{eq:L_alpha_diff}}
\end{eqnarray}
From here we see that there exist such $c_k$ that:
\begin{eqnarray}{\label{eq:L_alpha_diff_2}}
&& L_{\alpha'}(s)=\frac{c_0}{s}+\sum_{k}\frac{c_k}{s+b_k}.
\end{eqnarray}
Then, taking the inversion of the Laplace transform, we have the
expression for $\alpha'(x)$:
\begin{eqnarray*}
&& {\alpha'}(x) = {c_0}+\sum_{k}{c_k}e^{-b_k x}.
\end{eqnarray*}
and as $\alpha(0)=0$, then for $\alpha(x)$ we have
\begin{eqnarray*}
&& \alpha(x)=c_0 x - \sum_k{\frac{c_k}{b_k}e^{-b_k
x}}+\sum_k\frac{c_k}{b_k}, \quad \alpha(0)=0.
\end{eqnarray*}
 Now let us find
$c_k$. To find $c_0$ let us multiply both parts of the equation
({\ref{eq:L_alpha_diff_2}}) by $s$ and find the value at the point
$s=0$. Then, we have that
\begin{eqnarray}{\label{eq:c0_L}}
&& c_0=L_{\alpha'}(s) s|_{s=0} =
{\frac{\prod_{i}\mu_i}{\prod_{i}b_i}}.
\end{eqnarray}
From (\ref{eq:b_i_find_eq}) we have the following:
\begin{eqnarray*}
&& \frac{\prod_{i}b_i}{\prod_{i}\mu_i}=\Psi(s)|_{s=0}=1- \lambda
\overline{n}\sum_{i}{\frac {p_i }{\mu_i}} = 1- \lambda
\overline{n}m = {1-\rho}.
\end{eqnarray*}
So, then
\begin{eqnarray*}c_0=\frac{1}{1-\rho}.
\end{eqnarray*}
Let us find the other coefficients $c_i$. We denote:
\begin{eqnarray*}
&& L_{\alpha'}^*(s)=\sum_{i}\frac{c_i}{s+b_i}, \\
\Longrightarrow &&L_{\alpha'}(s) =
\frac{c_0}{s}+\sum_{i}\frac{c_i}{s+b_i} =
\frac{c_0}{s}+L_{\alpha'}^*(s).
\end{eqnarray*} According to the system
({\ref{eq:L_alpha_diff_mult}}),
({\ref{eq:L_alpha_diff}}) we have the following:
\begin{eqnarray*}
&& L_j^*=\sum_{i}\frac{c_i}{\mu_j+b_i}, \quad j=1,...,N, \\
(\ref{eq:L_alpha_diff_mult}) \Longrightarrow &&
L_{\alpha'}^{*}(s)\frac{\prod_{i}(s+b_i)}{\prod_{i}(s+\mu_i)}
=\lambda \overline{n}\sum_{i}{\frac {p_i L_i^{*}}{s+\mu_i}}+
{b}\sum_{i}{\frac {p_i}{s+\mu_i}}, \\
\Longrightarrow && L_{\alpha'}^*(s) \frac{\prod_{i}(s+b_i)}
{\prod_{i}(s+\mu_i)} (s+\mu_q)|_{s=-\mu_q} =\lambda \overline{n}
\sum_{i}{\frac {p_i L_i^*}{s+\mu_i}}(s+\mu_q)|_{s=-\mu_q}+\\
&&\quad \quad \quad\quad \quad \quad \quad \quad \quad\quad \quad \quad \quad \quad \quad
+{b}\sum_{i}{\frac {p_i}{s+\mu_i}}(s+\mu_q)|_{s=-\mu_q}, \quad  q=1,...,N, \\
\Longrightarrow && \sum_{j}\frac{c_j} {b_j-\mu_q}
\frac{\prod_{i}(b_i-\mu_q)} {\prod_{i \neq q}(\mu_i-\mu_q)}
=\lambda \overline{n}{p_q L_q^*}+ b {p_q}, \quad q=1,...,N, \\
\Longrightarrow && \sum_{j}\frac{c_j}{b_j-\mu_q}
\frac{\prod_{i}(b_i-\mu_q)}{\prod_{i \neq q}(\mu_i-\mu_q)}
=\lambda \overline{n} p_q \sum_{j}\frac{c_j}{b_j+\mu_q}+ b{p_q},
\quad q=1,...,N.
\end{eqnarray*}
Let us notice that from (\ref{eq:b_i_find_eq}) we have the
following:
\begin{eqnarray*}
&& \frac{\prod_{i}(b_i-\mu_q)}{\prod_{i \neq q}(\mu_i-\mu_q)}=
\frac{\prod_{i}(s+b_i)}{\prod_{i}(s+\mu_i)}(s+\mu_q)|_{s=-\mu_q}=\Psi(s)(s+\mu_q)|_{s=-\mu_q}=\\
&& =\left(1- \lambda \overline{n}\sum_{i}{\frac { p_i
}{s+\mu_i}}\right)(s+\mu_q)|_{s=-\mu_q} = -\lambda
\overline{n}\,p_q, \quad  q=1,...,N,
\end{eqnarray*} then
\begin{eqnarray*}
\Longrightarrow && \sum_{j}\frac{c_j}{b_j-\mu_q} (-\lambda
\overline{n} \,p_q) =\lambda \overline{n}\,p_q
\sum_{j}\frac{c_j}{b_j+\mu_q}+
{b}{p_q} ,\quad q=1,...,N,  \\
\Longrightarrow && \sum_{j}\frac{c_j}{\mu_q-b_j}-
\sum_{j}\frac{c_j}{\mu_q+b_j}=
\frac{b}{\lambda \overline{n}}, \quad q=1,...,N,   \\
\Longrightarrow && \sum_{j}\frac{c_j b_j}{\mu_q^2-b_j^2}=
{\frac{b}{2\lambda \overline{n}}}, \quad q=1,...,N.
\end{eqnarray*}
So, $c_j$ are solutions of the following linear system:
\begin{equation}{\label{eq:sys_c_j}}
\begin{array}{l}
\displaystyle \sum_{j}\frac{c_j b_j}{\mu_q^2-b_j^2}=
{\frac{b}{2\lambda \overline{n}}},\quad q=1,...,N.
\end{array}
\end{equation}
If we denote
\begin{eqnarray*}
&& x_k = \frac{c_k b_k}{\frac{b}{2\lambda \overline{n}}}, \quad
k=1,...,N,
\end{eqnarray*}
the system (\ref{eq:sys_c_j}) will take the form
({\ref{eq:x_j_sys}}) and by Lemma~{\ref{lemma_Cauchy_sys}}  for
$c_j$ the final result is as follows:
\begin{eqnarray*}
&& c_k= {\frac{b}{2\lambda \overline{n}}}
\left({\frac{x_k}{b_k}}\right)= {\frac{b}{2\lambda
\overline{n}}}\left(\frac {\displaystyle\prod_{q}
{(\mu_q^2-b_k^2)}} {b_k \displaystyle\prod_{q \neq
k}{(b_q^2-b_k^2)} }\right), \quad k=1,...,N.
\end{eqnarray*} This completes the proof of
Theorem~{\ref{th_alpha_result}}.
\end{proof}
\begin{Corollary} {\label{corollary:alpha_concave}}
The expected conditional sojourn time function in the BPS system
with hyper-exponential job size distribution as in (\ref{eq:B(x)})
is a strictly concave function with respect to job sizes.
\end{Corollary}
\begin{proof}[\emph{\textbf{Proof.}}] The function
\begin{eqnarray*}
&& \alpha(x)=c_0 x - \sum_k{\frac{c_k}{b_k}e^{-b_k
x}}+\sum_k\frac{c_k}{b_k}, \quad \alpha(0)=0\end{eqnarray*} is a
strictly concave function if $\alpha''(x) < 0$.
\begin{eqnarray*}
\alpha''(x)=- \sum_k{{c_k}{b_k}e^{-b_k x}} < 0
\end{eqnarray*}
as $c_k > 0, \, b_k>0, \,\, k=1,...,N$, which follows from $b>0,
\overline{n}>0 $, Corollary~{\ref{crlr:x_k_positive}} and
Lemma~{\ref{lemma_Psi_solut}}.
\end{proof}
\begin{Corollary}
The expected sojourn time in the BPS system with hyper-exponential
job size distribution as in ({\ref{eq:B(x)}}) is given by
\begin{eqnarray}
&&\overline{T}^{BPS}=\frac{m}{1-\rho}+\sum_{i,j}{\frac{p_i
c_j}{\mu_i+b_j}}.
\end{eqnarray}
\end{Corollary}
\begin{proof}[\emph{\textbf{Proof.}}] As the expected sojourn time
$\overline{T}^{BPS}$ is given by
\begin{eqnarray*}
&&\overline{T}^{BPS}=\int_0^\infty \alpha'(x)\overline{B}(x)dx,
\end{eqnarray*}
then using ({\ref{eq:alpha(x)}}) we receive the statement of the
Corollary.
\end{proof}

\newpage
\section{The Analysis of the Two Level Processor Sharing Model}{\label{sec_TLPS}}

\subsection{Main definitions}
\label{subsec:defin}
{\mbox{}}\\
We study the Two Level Processor Sharing (TLPS) scheduling
discipline with the hyper-exponential job size distribution. The
model description is as follows.
\\\\
Jobs arrive to the system according to a Poisson process with rate
$\lambda$.
Let $\theta$ be a given threshold. The jobs in the system that
attained a service less than $\theta$ are assigned to the high
priority queue. If in addition there are jobs with attained
service greater than $\theta$, such a job is separated into two
parts. The first part of size $\theta$ is assigned to the high
priority queue and the second part of size $x-\theta$ waits in the
lower priority queue. The low priority queue is served when the
high priority queue is empty. Both queues are served according to
the Processor Sharing (PS) discipline.
\\\\
Let us denote the job size distribution by $F(x)$. By
${\overline{F}(x)=1-F(x)}$ we denote the complementary
distribution function. We consider the case, when $F(x)$ is a
hyper-exponential distribution function, namely
\begin{equation}{\label{eq:F(x)}}
F(x)=1-\sum_{i=1}^{N}{\widetilde{p_i} e^{-\mu_i x}},\quad
\sum_{i}{\widetilde{p_i}}=1, \,\, \widetilde{p_i}>0,\,\,
\mu_i>0,\quad i=1,...,N,\quad 1<N\leq \infty.
\end{equation}
The mean job size is given by $m={\int_{0}^{\infty} x dF(x)}$ and
the system load is $\rho=\lambda m$. We assume that the system is
stable ($\rho<1$) and is in steady state.

\subsection{The expected conditional sojourn time for the TLPS model}
\mbox{}\\
Let us denote by $\overline{T}^{TLPS}(x)$ the expected conditional
sojourn time in the TLPS system for a job of size $x$ and by
$\overline{T}(\theta)$ the expected sojourn time of the system.
\\\\
According to {\cite{kleinrock}} the expected conditional sojourn
time of the system is given by:

\begin{eqnarray*}
\overline{T}^{TLPS}(x)= \left\{
\begin{array}{l l}
{\displaystyle \frac{x}{1-\rho_{ \theta }}}, & x\in[0, \theta ], \\
{\displaystyle \frac {\overline{W}( \theta )+ \theta +\alpha(x-
\theta )}{1-\rho_{ \theta }}}, & x\in(\theta,\infty),
\end{array}\right.
\end{eqnarray*}
where $\rho_{\theta}$ is the utilization factor for the truncated
distribution ${ \rho_{ \theta }= \lambda\,{ \overline{X_{ \theta
}^1}}}$, the $n$-th moment of the distribution truncated at
$\theta$ is
\begin{equation*}
{\label{Xn}}{\overline{X_{\theta}^n}} = \int _{0}^{ \theta }\!n
y^{n-1} \overline{F}(y) dy,
\end{equation*}
${(\overline{W}( \theta )+ \theta)/{(1-\rho_{ \theta })}}$
expresses the time needed to reach the low priority queue. This
time consists of the time ${\theta/{(1-\rho_{ \theta })}}$ spent
in the high priority queue, where the flow is served up to the
threshold $\theta$, plus the time ${\overline{W}( \theta
)/{(1-\rho_{ \theta })}}$ which is spent waiting for the high
priority queue to empty. Here ${\overline{W}(\theta )={\frac
{\lambda\,{ \overline{X_{ \theta }^2}}} {2(1-\rho_{ \theta })
}}}$. The remaining term ${{\alpha(x- \theta )}/{(1-\rho_{ \theta
})}}$ is the time spent in the low priority queue.
\\\\
To find $\alpha(x)$ we use the interpretation of the lower
priority queue as a PS system with batch arrivals.
\\\\
Let us denote
$$\overline{F_{\theta}^i}=\widetilde{p_i} e^{-\mu_i \theta},
\quad i=1,...,N, $$
and prove the following Theorem:
\begin{Theorem}{\label{th:TLPS_alpha}}
In TLPS priority queue  with the hyper-exponential job size
distribution as in ({\ref{eq:F(x)}}):
\begin{eqnarray*}
&&\alpha(x)=c_0(\theta) x - \sum_k
{\frac{c_k(\theta)}{b_k(\theta)}e^{-b_k(\theta) x}}+\sum_k
\frac{c_k(\theta)}{b_k(\theta)}, \quad
\alpha(0)=0,{\label{eq:alpha(x)_TLPS}}
\end{eqnarray*}
where
\begin{eqnarray*}
&& c_0(\theta)=\frac{1-\rho_{\theta}}{1-\rho},\label{eq:c_0_theta}\\
&& c_k(\theta)= {\frac{b}{2 \lambda \overline{n}}}\left( \frac
{\displaystyle\prod_{q=1,...,N} {(\mu_q^2-b_k^2(\theta))}} {
b_k(\theta)\displaystyle\prod_{q \neq
k}{(b_q^2(\theta)-b_k^2(\theta))} }\right),\quad
k=1,...,N,{\label{eq:c_k_theta}}
\end{eqnarray*}
and $b_i(\theta)$ are roots of
\begin{eqnarray*}
&& 1- \frac{\lambda}{1-\rho_{\theta}} \sum_{i}{\frac {
\overline{F_{\theta}^i} }{s+\mu_i}}=0,
\end{eqnarray*}
and satisfy the following inequalities:
$$0<b_N(\theta)<\mu_N,\,\,\,\mu_{i+1}<b_i(\theta)<\mu_i,\quad i=1,...,N-1. $$
\end{Theorem}
\begin{proof}[\emph{\textbf{Proof.}}]
As was shown in \cite{kleinrock},
$\alpha'(x)=d\alpha/dx$ is the solution of the integral equation
we looked at in Section {\ref{sec_BPS}}
\begin{equation*}
{\alpha'(x)=\lambda \overline{n} \int_{0}^{\infty} \alpha'(y)
\overline{B}(x+y)dy + \lambda \overline{n} \int_{0}^{x} \alpha'(y)
\overline{B}(x-y)dy +b \overline{B}(x)+1}.
\end{equation*}
Here the average batch size is given by ${
\overline{n}={\frac{1-F( \theta )}{1-\rho_{ \theta }}}}$ and the
average number of jobs that arrive to the low priority queue in
addition to the tagged job is given by $ b={\frac{2 \lambda\,
\left( 1-F \left( { \theta } \right) \right)\left( \overline{W} (
\theta ) +{ \theta } \right) }{(1-\rho_{ \theta })}}$. The
complementary truncated distribution is $ {\overline{B}(x)}=
{\frac {1-F(x+\theta)}{1-F(\theta)}}$, then:
\begin{eqnarray*}
&& \overline{B}(x) = \frac{\overline{F}(x+\theta)}
{\overline{F}(\theta)} = \frac{1}{\overline{F}(\theta)}
{\sum_i{\widetilde{p_i}e^{-\mu_i (x+\theta)}}}= \sum_i \frac
{\widetilde{p_i}e^{-\mu_i \theta}} {\overline{F}(\theta)}
e^{-\mu_i x} = \sum_i {p_i} e^{-\mu_i x},\\
&& p_i= \frac{1}{\overline{F}(\theta)}\widetilde{p_i} e^{-\mu_i
\theta}= \frac{1}{\overline{F}(\theta)} \overline{F_{\theta}^i},
\quad i=1,...,N.\\
\end{eqnarray*}
We apply the results of Theorem~{\ref{th_alpha_result}} for the
TLPS model. To calculate $c_0$  we use ({\ref{eq:c0_L}}) and
(\ref{eq:b_i_find_eq}) and get the following
\begin{eqnarray*}
\frac{1}{c_0(\theta)}=\frac{\prod_{i}b_i(\theta)}{\prod_{i}\mu_i}=\Psi(s)|_{s=0}=1-
\lambda \overline{n}\sum_{i}{\frac {p_i }{\mu_i}} = 1-
\frac{\lambda}{1-\rho_{\theta}}\sum_{i}{\frac
{\overline{F_{\theta}^i} }{\mu_i}} = 1- \frac{\lambda
(m-\overline{X^1_{\theta}})}{1-\rho_{\theta}} =
\frac{1-\rho}{1-\rho_{\theta}}.
\end{eqnarray*}
So, then
\begin{eqnarray*}
c_0(\theta)=\frac{1-\rho_{\theta}}{1-\rho}.
\end{eqnarray*}
For $c_k(\theta), b_k(\theta)$ we use the results of
Theorem~{\ref{th_alpha_result}} and get the statement of
Theorem~{\ref{th:TLPS_alpha}}.
\end{proof}
\begin{Corollary} {\label{corolary:c_k_theta_positive}} The coefficients
$c_k(\theta)>0, \quad k=1,...,N$ if $\theta>0$.
\end{Corollary}
\begin{proof}[\emph{\textbf{Proof.}}] As ({\ref{eq:mu_cond}}), \mbox{$
0<b_N(\theta)<\mu_N, \,\,\,\mu_{i+1}<b_i(\theta)<\mu_i, \quad
i=1,...,N-1$} and \mbox{$ b>0 $}, \mbox{$ \overline{n}>0 $} when
\mbox{$ \theta >0 $}, then the statement of Corollary holds.
\end{proof}
\begin{Corollary}
For the TLPS queue with hyper-exponential job size distribution as
in ({\ref{eq:F(x)}}) the function $\alpha(x)$ is a strictly
concave function with respect to job sizes with positive values of
$\theta$.
\end{Corollary}
\begin{proof}[\emph{\textbf{Proof.}}] The function
\begin{eqnarray*}
&&\alpha(x)=c_0(\theta) x -
\sum_k{\frac{c_k(\theta)}{b_k(\theta)}e^{-b_k(\theta)
x}}+\sum_k\frac{c_k(\theta)}{b_k(\theta)}, \quad
\alpha(0)=0\end{eqnarray*} is a strictly concave function if
$\alpha''(x)<0$.
$$\alpha''(x)=- \sum_k{{c_k(\theta)}{b_k(\theta)}e^{-b_k(\theta) x}}<0$$
as $c_k(\theta)>0, b_k(\theta)>0,\,\,\, k=1,...,N, $ if $\theta>0$
as it follows from Corollary~{\ref{corolary:c_k_theta_positive}}
and Theorem~{\ref{th:TLPS_alpha}}.
\end{proof}

\begin{Corollary}
The expected conditional sojourn time for the TLPS queue with
hyper-exponential job size distribution as in ({\ref{eq:F(x)}}) is
not a concave function with respect to job sizes.
\end{Corollary}
\begin{proof}[\emph{\textbf{Proof.}}] The function
${\overline{T}^{TLPS}(x)}$ is a concave function, when
\begin{eqnarray*}
&& \overline{T}^{TLPS}\!\,'(x)|_{x=\theta-0} \geq
\overline{T}^{TLPS} \!\,'(x)|_{x=\theta+0} \\
&& \frac{1}{1-\rho_{ \theta }} |_{x=\theta-0} \geq {\displaystyle
\frac {\alpha'(x- \theta )}{1-\rho_{ \theta }}} |_{x=\theta+0}\\
&& 1 \geq {\alpha'(0+ )}.
\end{eqnarray*}
As shown in {\cite{AAB_BPS_MLPS_QS}},
\begin{eqnarray*}
&& \alpha(x)\geq\frac{x}{1-\rho},\\
&& \alpha'(x)\geq \frac{1}{1-\rho}> 1.
\end{eqnarray*}
Then, it follows that ${\overline{T}^{TLPS}(x)}$ is not a concave
function.
\end{proof}

\subsection{The expected sojourn time for the TLPS model}
\mbox{}
\begin{Theorem} The expected sojourn time in the TLPS system with
hyper-exponential distribution function ({\ref{eq:F(x)}}) is given
by the following equation:
\begin{eqnarray*}
&&{\overline{T}(\theta)}= \frac{{\overline{X^1_{\theta}}}
+\overline{W}(\theta)\overline{F}(\theta)}{1-\rho_{\theta}} +
{\frac{(m-{\overline{X^1_{ \theta}}})}{1-\rho}}
+{\frac{(\overline{W}(\theta)+\theta)}{1-\rho_{\theta}}}\sum_{i,j}\frac{
\overline{F_{\theta}^i}}{b_j(\theta)(\mu_i+b_j(\theta))}\frac
{\displaystyle\prod_{q} {(\mu_q^2-b_j^2(\theta))}} {
\displaystyle\prod_{q \neq j}{(b_q^2(\theta)-b_j^2(\theta))} },
\end{eqnarray*}
where $b_i(\theta)$ are defined as in
Theorem~{\ref{th:TLPS_alpha}}.
\end{Theorem}
\begin{proof}[\emph{\textbf{Proof.}}] According to {\cite{kleinrock}} and
Theorem~{\ref{th:TLPS_alpha}} we have the following:
\begin{eqnarray*}
&&\overline{T}(\theta)=
\frac{{\overline{X_{\theta}^1}}+\overline{W}( \theta
)\overline{F}( \theta )}{1-\rho_{ \theta }} +
{\frac{1}{1-\rho_{\theta}}}\overline{T}^{BPS}(\theta),\nonumber\\
{\label{T_BPS}} &&\overline{T}^{BPS}(\theta) = {\int_{ \theta
}^{\infty}{{\alpha(x- \theta )}}dF(x)=\int_0^\infty
\alpha'(x)\overline{F}(x+\theta)dx},\\
(Theorem~{\ref{th:TLPS_alpha}}) \Longrightarrow &&
\overline{T}^{BPS}(\theta)=c_0(\theta)
(m-{\overline{X^1_{\theta}}})+\sum_{i,j}\frac{
\overline{F_{\theta}^i} c_j(\theta)}{\mu_i+b_j(\theta)},
\end{eqnarray*}
then
\begin{eqnarray*}
&&{\overline{T}(\theta)}= \frac{{\overline{X^1_{
\theta}}}+\overline{W}( \theta )\overline{F}( \theta )}{1-\rho_{
\theta }} + {\frac{(m-{\overline{X^1_{ \theta}}})}{1-\rho}}
+{\frac{(\overline{W}(\theta)+\theta)}{1-\rho_{\theta}}}\sum_{i,j}\frac{
\overline{F_{\theta}^i}}{b_j(\theta)(\mu_i+b_j(\theta))}\frac
{\displaystyle\prod_{q} {(\mu_q^2-b_j^2(\theta))}} {
\displaystyle\prod_{q \neq j}{(b_q^2(\theta)-b_j^2(\theta))} }.
\end{eqnarray*}
\end{proof}

\newpage
\section{Conclusion}

We study the BPS queueing model, when the job size distribution is
hyper-exponential, and we find an analytical expression of the
expected conditional response time and for the expected sojourn
time. We show that the function of the expected conditional
sojourn time in the BPS system with hyper-exponential job size
distribution is a concave function with respect to job sizes. We
apply the results obtained for the BPS model to the TLPS
scheduling mechanism with the hyper-exponential job size
distribution and we find the expressions of the expected
conditional response time and expected response time for the TLPS
model.

\section*{Acknowledgment}

I would like to thank K. Avrachenkov and P. Brown for fruitful
discussions and suggestions.

\newpage
\section{Appendix}\mbox{}
\textbf{Lemma \,{\ref{lemma_Cauchy_sys}}.} The solution of the
following system of linear equations
\begin{equation}{\label{eq:x_j_sys_2}}
\begin{array}{l}
\displaystyle \sum_{j}\frac{x_j }{\mu_q^2-b_j^2}= 1,\quad
q=1,...,N,
\end{array}
\end{equation}
is given by:
\begin{equation}
x_k= \frac {\displaystyle\prod_{q=1,...,N} {(\mu_q^2-b_k^2)}} {
\displaystyle\prod_{q \neq k}{(b_q^2-b_k^2)} }.
\end{equation}
\begin{proof}[\emph{\textbf{Proof.}}] Let \begin{eqnarray*}
&&x=[x_1, x_2,...,x_N]^T, \\
&&\underline{1}=[1,1,...,1]^T |_{1\times N}
\end{eqnarray*}
be two vectors of size $N$,
\begin{equation*}
D=\left(
\begin{array}{l l l l }
\frac{1 }{\mu_1^2-b_1^2}& \frac{1 }{\mu_1^2-b_2^2}&...& \frac{1 }{\mu_1^2-b_N^2}\\
\frac{1 }{\mu_2^2-b_1^2}& \frac{1 }{\mu_2^2-b_2^2}&...& \frac{1 }{\mu_2^2-b_N^2}\\
.....\\
\frac{1 }{\mu_N^2-b_1^2}& \frac{1 }{\mu_N^2-b_2^2}&...& \frac{1 }{\mu_N^2-b_N^2}\\
\end{array}\right)_{n \times N}
\end{equation*}
be the matrix of size $N\times N$. Then equation
({\ref{eq:x_j_sys_2}}) could be rewritten as
\begin{eqnarray*}&&D x = \underline{1}.
\end{eqnarray*}
Applying the Cramer formulas {\cite{Kurosh}} we obtain:
\begin{eqnarray}
&&x_k =\frac{\det D_k}{\det D}=
\frac{\det{[D_{[1]},...D_{[k-1]},\underline{1},D_{[k+1]},...D_{[N]}]}}{\det{D}},
\quad k=1,...,N, {\label{eq:x_k_det}} \\
&&D_k
=[D_{[1]},...D_{[k-1]},\underline{1},D_{[k+1]},...D_{[N]}].\nonumber
\end{eqnarray}
Since $D$ is a Cauchy matrix, its determinant is known
{\cite{Kurosh}}:
\begin{equation}{\label{eq:det_D}}
\displaystyle \det{D} = \frac{\displaystyle\prod_{1\leq j< k \leq
N}{(\mu_j^2-\mu_k^2)(b_k^2-b_j^2)}}
{\displaystyle\prod_{j,k=1,...,N}{(\mu_j^2-b_k^2)} }\,\,.
\end{equation}
As  \mbox{ $ 0 < b_N < \mu_N <
...<\mu_{i+1}<b_i<\mu_i<...<b_1<\mu_1 $ }, then \mbox{$ \det{D}>0
$} and we can use Cramer formulas to calculate $x_k$. Let us find
\mbox{$\det{D_k}=\det{[D_{[1]},...D_{[k-1]},\underline{1},D_{[k+1]},...D_{[N]}]}$}.
\begin{equation*}D_k=\left(
\begin{array}{l l l l l l l}
\frac{1 }{\mu_1^2-b_1^2}&...& \frac{1 }{\mu_1^2-b_{k-1}^2} & 1 & \frac{1 }{\mu_1^2-b_{k+1}^2}  &...& \frac{1 }{\mu_1^2-b_N^2}\\
\frac{1 }{\mu_2^2-b_1^2}&...& \frac{1 }{\mu_2^2-b_{k-1}^2} & 1 & \frac{1 }{\mu_2^2-b_{k+1}^2}  &...& \frac{1 }{\mu_2^2-b_N^2}\\
.....\\
\frac{1 }{\mu_N^2-b_1^2}&...& \frac{1 }{\mu_N^2-b_{k-1}^2} & 1 & \frac{1 }{\mu_N^2-b_{k+1}^2}  &...& \frac{1 }{\mu_N^2-b_N^2}\\
\end{array}\right)_{{N \times N}}
\end{equation*}
\begin{equation*}\det{D_k}=\left|
\begin{array}{l l l l l l l}
\frac{1 }{\mu_1^2-b_1^2}&...& \frac{1 }{\mu_1^2-b_{k-1}^2} & 1 & \frac{1 }{\mu_1^2-b_{k+1}^2}  &...& \frac{1 }{\mu_1^2-b_N^2}\\
\frac{1 }{\mu_2^2-b_1^2}&...& \frac{1 }{\mu_2^2-b_{k-1}^2} & 1 & \frac{1 }{\mu_2^2-b_{k+1}^2}  &...& \frac{1 }{\mu_2^2-b_N^2}\\
.....\\
\frac{1 }{\mu_N^2-b_1^2}&...& \frac{1 }{\mu_N^2-b_{k-1}^2} & 1 & \frac{1 }{\mu_N^2-b_{k+1}^2}  &...& \frac{1 }{\mu_N^2-b_N^2}\\
\end{array}\right|_{N\times N}
\end{equation*}
\begin{equation*}=(-1)^{k-1}\left|
\begin{array}{l l l l l l l}
1 & \frac{1 }{\mu_1^2-b_1^2}&...& \frac{1 }{\mu_1^2-b_{k-1}^2} &  \frac{1 }{\mu_1^2-b_{k+1}^2}  &...& \frac{1 }{\mu_1^2-b_N^2}\\
1 & \frac{1 }{\mu_2^2-b_1^2}&...& \frac{1 }{\mu_2^2-b_{k-1}^2} &  \frac{1 }{\mu_2^2-b_{k+1}^2}  &...& \frac{1 }{\mu_2^2-b_N^2}\\
.....\\
1 & \frac{1 }{\mu_N^2-b_1^2}&...& \frac{1 }{\mu_N^2-b_{k-1}^2} &  \frac{1 }{\mu_N^2-b_{k+1}^2}  &...& \frac{1 }{\mu_N^2-b_N^2}\\
\end{array}\right|_{N \times N}
\end{equation*}
\\
To simplify the following computations let us introduce the
following notations:
\begin{eqnarray*}
&&\beta_i=-b_{i-1}^2, \quad i=2,...,k,\\
&&\beta_i=-b_{i}^2,\quad i=k+1,...,N.
\end{eqnarray*}
Then, we have
\begin{equation*}\det{D_k}=(-1)^{k-1}\left|
\begin{array}{l l l l l l l}
1 & \frac{1 }{\mu_1^2+\beta_2}&...& \frac{1 }{\mu_1^2+\beta_k} &  \frac{1 }{\mu_1^2+\beta_{k+1}}  &...& \frac{1 }{\mu_1^2+\beta_N}\\
1 & \frac{1 }{\mu_2^2+\beta_2}&...& \frac{1 }{\mu_2^2+\beta_k} &  \frac{1 }{\mu_2^2+\beta_{k+1}}  &...& \frac{1 }{\mu_2^2+\beta_N}\\
.....\\
1 & \frac{1 }{\mu_N^2+\beta_2^2}&...& \frac{1 }{\mu_N^2+\beta_{k}} &  \frac{1 }{\mu_N^2+\beta_{k+1}}  &...& \frac{1 }{\mu_N^2+\beta_N}\\
\end{array}\right|_{N \times N}
\end{equation*}
\begin{equation*}\det{D_k}=(-1)^{k-1}\left|
\begin{array}{l l l  l l l}
1 & \frac{1 }{\mu_1^2+\beta_2}&...& \frac{1 }{\mu_1^2+\beta_k} &...& \frac{1 }{\mu_1^2+\beta_N}\\
0 & \frac{\mu_1^2-\mu_2^2}{(\mu_2^2+\beta_2)(\mu_1^2+\beta_2)}&...& \frac{\mu_1^2-\mu_2^2 }{(\mu_2^2+\beta_k)(\mu_1^2+\beta_k)} &...& \frac{\mu_1^2-\mu_2^2 }{(\mu_2^2+\beta_N)(\mu_1^2+\beta_N)}\\
.....\\
0 & \frac{\mu_1^2-\mu_N^2 }{(\mu_N^2+\beta_2)(\mu_1^2+\beta_2)}&...& \frac{\mu_1^2-\mu_N^2 }{(\mu_N^2+\beta_{k})(\mu_1^2+\beta_k)} &...& \frac{\mu_1^2-\mu_N^2 }{(\mu_N^2+\beta_N)(\mu_1^2+\beta_N)}\\
\end{array}\right|_{N \times N}
\end{equation*}
\begin{equation*}=(-1)^{k-1}\frac{(\mu_1^2-\mu_2^2)...(\mu_1^2-\mu_N^2)}{(\mu_1^2+\beta_2)...(\mu_1^2+\beta_N)}\left|
\begin{array}{l l l l l l}
\frac{1 }{\mu_2^2+\beta_2}&...& \frac{1 }{\mu_2^2+\beta_k} &  \frac{1 }{\mu_2^2+\beta_{k+1}}  &...& \frac{1 }{\mu_2^2+\beta_N}\\
.....\\
\frac{1 }{\mu_N^2+\beta_2^2}&...& \frac{1 }{\mu_N^2+\beta_{k}} &  \frac{1 }{\mu_N^2+\beta_{k+1}}  &...& \frac{1 }{\mu_N^2+\beta_N}\\
\end{array}\right|_{(N-1) \times (N-1)}
\end{equation*}
So, as the above matrix under the sign of determinant is a Cauchy
matrix of size $N-1$, the following equation holds:
\begin{equation*}\det{D_k}=(-1)^{k-1}\frac{(\mu_1^2-\mu_2^2)...(\mu_1^2-\mu_N^2)}{(\mu_1^2+\beta_2)...(\mu_1^2+\beta_N)}
\frac{\displaystyle\prod_{2\leq j< q \leq
N}{(\mu_j^2-\mu_q^2)(\beta_j-\beta_q)}}
{\displaystyle\prod_{j,q=2,...,N}{(\mu_j^2+\beta_q)} }\,\,.
\end{equation*}
Let us recall that $\beta_i=-b_{i-1}^2\quad i=2,k,\quad
\beta_i=-b_{i}^2\quad i=k+1,...,N$, then
\begin{eqnarray*}
\det{D_k} &=& (-1)^{k-1}
\frac{\displaystyle\prod_{q=2,...,N}(\mu_1^2-\mu_q^2)}
{\displaystyle\prod_{q=2,...,N}(\mu_1^2+\beta_q)}
\frac{\displaystyle\prod_{2\leq j< q \leq
N}{(\mu_j^2-\mu_q^2)(\beta_j-\beta_q)}}
{\displaystyle\prod_{j,q=2,...,N}{(\mu_j^2+\beta_q)} }\\
& =& (-1)^{k-1} \frac{\displaystyle \prod_{1\leq j< q \leq
N}{(\mu_j^2-\mu_q^2) \prod_{2\leq j< q \leq
N}{(\beta_j-\beta_q)}}} {\displaystyle\prod_{j=1,...,N
,q=2,...,N}{(\mu_j^2+\beta_q)} } \\
& =&(-1)^{k-1} \frac{\displaystyle\prod_{1\leq j<q \leq
N}{(\mu_j^2-\mu_q^2)\prod_{1\leq j<q \leq N,\, j,q \neq
k}(b_q^2-b_j^2)}} {\displaystyle\prod_{j,q=1,...,N, q \neq
k}{(\mu_j^2-b_q^2)} }\\
& =&(-1)^{k-1} \frac{\displaystyle\prod_{1\leq j < q  \leq N}
{(\mu_j^2-\mu_q^2)\prod_{1\leq j < q \leq N}(b_q^2-b_j^2)}}
{\displaystyle\prod_{j,q=1,...,N, q \neq k}{(\mu_j^2-b_q^2)}
(-1)^{k-1}\displaystyle\prod_{q=1,...,N,\,q \neq
k}{(b_q^2-b_k^2)}} \\
& =& \frac{\displaystyle\prod_{1\leq j < q  \leq N}
{(\mu_j^2-\mu_q^2)(b_q^2-b_j^2)}\displaystyle\prod_{j=1,...,N}
{(\mu_j^2-b_k^2)}} {\displaystyle\prod_{j,q=1,...,N}
{(\mu_j^2-b_q^2)} \displaystyle \prod_{q=1,...,N, \,q \neq k}
{(b_q^2-b_k^2)}} .
\end{eqnarray*}
Then, from (\ref{eq:x_k_det}) and (\ref{eq:det_D}), we have the
following expressions for $x_k$:
\begin{equation*}
x_k= \frac {\displaystyle\prod_{q=1,...,N} {(\mu_q^2-b_k^2)}}
 {\displaystyle \prod_{q=1,...,N,\, q\neq k}{(b_q^2-b_k^2)}},
\end{equation*}
what proves Lemma~{\ref{lemma_Cauchy_sys}}.
\end{proof}

\newpage
\tableofcontents

\end{document}